\documentclass[prl,aps,subeqns.floatfix,amsmath]{revtex4}
\usepackage[pdftex]{graphicx}
\usepackage{amsmath}
\usepackage{amsfonts}
\usepackage{epstopdf}
\usepackage{graphicx}
\usepackage{subfigure}
\usepackage{hyperref}
\usepackage{color}
\newtheorem{thm}{Theorem}
\begin{document}
\title{Self-adjoint extension for Maxwell-Chern-Simons model in long wavelength limit }
\author{Pinaki Patra}
\email{monk.ju@gmail.com}
\affiliation{Dept. of Physics, Brahmananda Keshab Chandra College, Kolkata, India-700108}
\affiliation{Dept. of Physics, University of Kalyani, West Bengal, India-741235}

\author{Kalpana Biswas}
\email{klpnbsws@gmail.com}
\affiliation{Dept. of Physics, University of Kalyani, West Bengal, India-741235}
\author{Jyoti Prasad Saha}
\email{jyotiprasadsaha@gmail.com}
\affiliation{Dept. of Physics, University of Kalyani, West Bengal, India-741235}

\date{\today}

\begin{abstract}
In the long wavelength limit, Maxwell-Chern-Simmon model and the dynamics of a particle in a plane under an external
magnetic field perpendicular to that plane are identical.  The self adjoint extension of such a problem depends
on the value of angular momentum quantum number $l$. In this article, we have shown that for $l\neq 0$, the
operator describing the Landau level wave-function is self adjoint; whereas, for $l=0$, infinite number of self-adjoint extension by an one parameter 
unitary mapping is possible. 
\end{abstract}
 
 \maketitle
\section{Introduction }
In the last few years there has been an increasing interest in the physics of quantum systems 
confined in a lower dimension (e.g, $1+1, 1+2$) and in the prominent role of quantum boundary 
conditions. Planar physics- physics in two spatial dimension-presents many interesting
surprises \cite{Gerald}. For example, there exist a new type of gauge theory- different
from Maxwell theory in $2+1$ dimension- namely, Chern-Simons theory. The long wavelength
limit (in which, we drop the all spatial derivatives) of Maxwell-Chern-Simons (MCS) Lagrangian \cite{Dune, Kogan} 
\begin{equation}
L = \frac{1}{2e^2}\dot{A}_i\dot{A}_i + \frac{\kappa}{2} \epsilon^{ij} \dot{A}_i A_j\;\;;\;i=1,2
\end{equation} 
has the same form as that of Lagrangian of a nonrelativistic charged particle moving in the
plane in presence of uniform magnetic field $b$ perpendicular to the plane 
\begin{equation}
L = \frac{m}{2}\dot{x}_i\dot{x}_i + \frac{b}{2} \epsilon^{ij} \dot{x}_i x_j
\end{equation}
Where, $\epsilon_{ij}$ is completely anti-symmetric Levi-ci-Vita symbol. The Hamiltonian is
\begin{equation}
H = \frac{1}{2} m v_i^2 \;\;,\; \mbox{where,} \;\; v_i = \frac{1}{m} (p_i - \frac{b}{2} \epsilon^{ij} x_j)
\end{equation}
Canonical quantization $[x_i, p_j]=i\delta_{ij}$ (where, $\delta_{ij}$ is Kronecker delta) 
implies $[v_i,v_j]=-i\frac{b}{m^2}\epsilon_{ij}$. So, the Therefore, MCS theory
and Landau problem \cite{Ashok} has direct correspondence; we have to just identify the proper
couplings e.g, $\kappa \rightleftharpoons b$.
Another way to express Landau level wave-functions 
\cite{Gerald, Landau, Aharonov, Dune, Kogan, Carlip, Zhang, Pisarski, Deser} is to express the Hamiltonian in the form 
\begin{eqnarray}
H=-\frac{1}{2m}(D_1^2+D_2^2) \\
\mbox{where,}\; D_1=\partial_1 + i\frac{b}{2}x_2, \;\;  D_2=\partial_2 - i\frac{b}{2}x_1
\end{eqnarray} 
under the following boundary condition (b.c)
\begin{equation}
\psi(z)=0 \;\;\;, \;\; for \mod{z}\rightarrow \infty
\end{equation}
Using factorization method \cite{Jana, Mielnik, Fernandez,Finkel,David, Shi, Pinaki, Ikot}, 
if we split the Hamiltonian as $-\frac{1}{2m}D_-D_+ + \frac{b}{2m}$ (where, $D_\pm=D_1\pm iD_2$), 
one can obtain the solution under this b.c.
\begin{equation}
\psi(z)= f(z)e^{-\frac{b}{4}{|z|^2}}\;\; \mbox{for \;\ some \\; holomorphic \;\; f}
\end{equation}
However, the same symmetric operator can be made self-adjoint under a class of b.c under which 
the well known studied above-mentioned b.c is included. In practical purpose, we can only regulate
the boundary conditions and can study and verify different behavior of  the quantum mechanical
system experimentally. Therefore, it is always interesting to find out possible choices of boundary
conditions under which a symmetric operator can be extended to a self-adjoint operator \cite{Stone,Von Neumann,Vanilse, Jordan,Guy}. \\
Extension problem \cite{Stone,Von Neumann,Vanilse, Jordan, Guy, Paolo} is not new and date back to 1929. But,
it is still interesting mathematically as well as to serve the purpose of Physics due to its recent
relevancy in various field.\\
In this article, we have studied the self-adjoint extension problem for the symmetric operator (1) which
is commonly popular to describe the Landau level wave function inside a graphene.

\section{Self adjoint extension of MCS model in long wavelength limit }
Consider the Hermitian operator 
\begin{equation}
	H=-\dfrac{1}{2m}(D_{1}^{2}+D_{2}^{2})
\end{equation}\\
Where, 
\begin{equation}
	D_{1}=\dfrac{\partial}{\partial x_{1}}
	+i\dfrac{b}{2}x_{2} \;\; \mbox{and,}\;\;
	D_{2}=\dfrac{\partial}{\partial x_{2}}
	-i\dfrac{b}{2}x_{1}
\end{equation}
Clearly, this is a symmetric operator. We can see that the case for $b=0$ is just the Laplacian
operator which is well understood. The case of $b\neq 0$ i.e, the presence of external magnetic
field makes the problem non-trivial. The appearance of two co-ordinate variables $x_1$ and $x_2$ suggests
that we should use polar co-ordinates to study the operator further. Therefore, let us consider
\begin{equation}
	x_{1}=r\cos\theta, \;\;\;
	x_{2}=r\sin\theta
\end{equation}
Now $H$ reads
\begin{equation}
	H = - \frac{1}{2m}
	\left[ \frac{\partial^{2}}{\partial r^{2}} + \frac{1}{r}
	\frac{\partial}{\partial r}
	+ \frac{1}{r^{2}}\frac{\partial^{2}}{\partial\theta^{2}}
 	-ib \frac{\partial}{\partial\theta } - \frac{b^2r^2}{4} \right] \end{equation}
Separation of variable method always makes life simple. But, to prepare $H$ for separation of variable
we have to notice that \cite{Cesar},
\begin{equation}
	\mbox{domain} (\hat{H})=C_{0}^{\infty}(\{\mathbb{R}^2\}/{0})
\end{equation}
Where, $C_{0}^{\infty}(\{\mathbb{R}^2\}/{0})$ is the space of compactly supported infinitely differentiable
functions over the complex field $\mathcal{C}$ and zero does not belong to the support.
As, the polar co-ordinate $r\in (0,\infty)$ and the angular variable $\theta \in (0,2\pi)$, we can split by
\begin{equation}
	L^{2}(\{\mathbb{R}^2\})=L^{2}(0,\infty),rdr)\otimes L^{2}(S^{1}d\theta)
\end{equation}

Let
\begin{equation}
	\mathcal{D}=span\{f(r)g(\theta),\; f\in C^{\infty}_{0}(0,\infty),\;g(\theta)\in C^{\infty}_{0}(S^{1})\} 
\end{equation}
Where, $S^1$ is the unit circle in $\mathbb{R}^2$.
Further let $\mathcal{D}$ is dense in $L^{2}(\mathcal{C})$.
Now, if we apply $\hat{H}$ on the functions $\psi(r,\theta)=f(r)g(\theta)$, we get
\begin{equation}
	\hat{H}f(r)g(\theta)=\left[ \left( \dfrac{\partial^{2}}{\partial r^{2}}+\dfrac{1}{r}
	\dfrac{\partial}{\partial r}\right) f(r)-\dfrac{b^{2}r^{2}}{4}f(r)\right] g(\theta) +
	\dfrac{f(r)}{r^{2}}Bg(\theta)+bf(r)B_{0}g(\theta)
\end{equation}
Where, 
\begin{equation}
	B_{0}=-i\dfrac{\partial}{\partial\theta} , \;\;\mbox{and,}\;\;
	B=\dfrac{\partial^{2}}{\partial\theta^{2}}
\end{equation}
The Laplace Beltrami Operator $B$ is self-adjoint on $L^{2}(S^{1},d\theta)$
and has eigenvectors 
\begin{equation} 
g_{l}(\theta) = \dfrac{1}{\sqrt{2}\pi}e^{il\theta}
\end{equation} 
with the eigenvalue $-l^{2},\; l\in\mathbb{Z}$, where $\mathbb{Z}$ is the set of integers.
Clearly, $\{g_{l}(\theta)\}_l$  constitutes an orthonormal basis of $L^{2}(S^{1},d\theta)$.\\
Further, $B_{0}$ is  self-adjoint in $L^{2}(S^{1},d\theta)$ with eigenvectors
 \begin{equation}
h_{n}(\theta)=\dfrac{1}{\sqrt{2}\pi}e^{in\theta} \;,\;n\in\mathbb{Z}
\end{equation}
 with the eigenvalues $n\in\mathbb{Z}$.\\
 If,$\mathcal{E}_{B}$ and
  $\mathcal{E}_{B_0}$ are eigen-subspaces of $B$ and $B_{0}$ respectively,
  then  $\mathcal{E}_{B}\subseteq \mathcal{E}_{B_{0}}$. 
  Let ,$\left[g_{l}\right]= \mbox{span}\left\lbrace g_{l}(\theta)\right\rbrace =\mathcal{E}_{B}$ ;
  $l\in\mathbb{Z}$, and 
  $\left[h_{n}\right]=\mbox{span} \left\lbrace h_{n}(\theta)\right\rbrace =\mathcal{E}_{B_0}\; ;
  n\in\mathbb{Z}$.
   Though it seems  $\mathcal{E}_{B}\subseteq \mathcal{E}_{B_{0}}$, but because
  of the fact $\mathbb{Z}$ and $\mathcal{Z}=\{n^2; n\in \mathbb{Z}\}$ are isomorphic to each other,
  it is not difficult to understand that  $\left[g_{l}\right]$ and  $\left[h_{n}\right]$  are isomorphic
  to each other.
 
  Let , \begin{eqnarray}
  L_{l}=L^{2}\left( (0,\infty), rdr\right) \otimes\left[g_{l}\right]\otimes\left[h_{l}\right]\\
  \mbox{Therefore,}\;\; L^{2}\left( \mathbb{R}^{2}\right)=\bigoplus L_{l} \;\;,\;\; l\in\mathbb{Z}
  \end{eqnarray}
       
  If $I_{l}$ is the identity operator of $\left[g_{l}\right]$ and $\left[h_{n}\right]$ , the restriction of
  $\hat{H}$ to $\mathcal{D}_{l} = \mathcal{D}\cap L_{l}$ is given by
  $\hat{H}| _{\mathcal{D}_{l}}=\hat{H_{l}}\otimes\hat{I_{l}}$ ; with
  \begin{equation}
  \hat{H_{l}}=\left( \dfrac{\partial^{2}}{\partial r^{2}}+\dfrac{1}{r}
  \dfrac{\partial}{\partial r}\right) -\dfrac{b^{2}r^{2}}{4}-\dfrac{l^{2}}{r^{2}}
  \end{equation}
  with domain $C^{\infty}_{0}(0,\infty)$.
  Now the question is whether we can find the self-adjoint extension of such restrictions.
  
  Considering the following unitary operator $U$
  \begin{eqnarray}
  U:L_{2}((0,\infty);r dr)\longrightarrow L_{2}((0,\infty);dr)(U\phi)(r)\\
  (U\phi)(r)=\sqrt{r}\phi(r),
  \end{eqnarray}
  Under this transformation $\hat{H}$ reduces to
 \begin{equation}
 \hat{\mathcal{H}_{l}}=	U\hat{H_{l}}U^{-1}=\dfrac{\partial^{2}}{\partial r^{2}}+
 (\dfrac{1}{4}-l^{2})\dfrac{1}{r^{2}}-\dfrac{b^{2}r^{2}}{4}
 	\end{equation}
Thus we had absorbed the term  	$\dfrac{\partial}{\partial r}$.
 	
But, being U is unitary, $Dom(\hat{\mathcal{H}_{l}})= C^{\infty}_{0}(0,\infty)$ ,  \\
$\hat{\mathcal{H}_{l}}^{\dagger}$ has the same action as $\hat{\mathcal{H}_{l}}$ 
but with different domain. In particular,
\begin{equation}
 dom \hat{\mathcal{H}_{l}}^{\dagger}=\left\lbrace \phi \in L^{2}(0,\infty)	
 : \phi , \phi^{'}\in A^{c}(0,\infty) , \hat{H_{l}}^{\dagger}\phi\in L^{2}(0,\infty)\right\rbrace .
\end{equation}
An operator $H$ is said to be self adjoint if $\mbox{dom}(H) = \mbox{dom}(H^\dagger)$
Therefore, to make $\hat{H}$ self-adjoint, we have to solve the extension problem. To do so, we have to identify the deficiency indices $n_+,\; n_- $ by
\begin{eqnarray}
n_+ = \mbox{dimension\; of}\; \mathcal{N}_+ \; ; \; \mathcal{N}_+ = \{ \psi \in \mbox{dom}(H^\dagger), \; H^\dagger \psi = z_+ \psi, \; Im(z_+) > 0 \} \nonumber \\
n_- = \mbox{dimension\; of}\; \mathcal{N}_- \; ; \; \mathcal{N}_- = \{ \psi \in \mbox{dom}(H^\dagger), \; H^\dagger \psi = z_- \psi, \; Im(z_-) < 0 \} \nonumber
\end{eqnarray}
Actually, it is sufficient to take $z_\pm = \pm i$. The necessary and sufficient condition to be the operator self adjoint is $n_+=n_-=0$. If, $n_+\neq n_-$, no self-adjoint extension is possible. However, for $n_+ = n_- =n \geq 1$, $H$ has infinitely many self-adjoint extensions parametrized by a $n\times n$ unitary matrix\cite{Von Neumann, Weyl}. \\
Therefore, we have to first consider the equations
\begin{equation}
\hat{\mathcal{H}}\psi_{\pm}=\pm i \psi_{\pm}
\end{equation}
and find out the number of independent solutions. To do so, we write down the equations explicitly. In particular,
\begin{eqnarray}
-\frac{1}{2m}\left[ \dfrac{d^{2}}{dr^{2}}+(\dfrac{1}{4}-l^{2})\dfrac{1}{r^{2}}-\dfrac{b^{2}}{4}r^{2}\right] 
\psi_{+}=i\psi_{+}\\
-\frac{1}{2m}\left[ \dfrac{d^{2}}{dr^{2}}+(\dfrac{1}{4}-l^{2})\dfrac{1}{r^{2}}-\dfrac{b^{2}}{4}r^{2}\right] 
\psi_{-}=-i\psi_{-}
\end{eqnarray}
 This equation(27) has two linearly independent solutions 
 \begin{eqnarray}
 \psi_{1+}(r)= \left(\frac{4}{b}\right)^\frac{l+1}{2} \frac{1}{\sqrt{r}} \mathcal{W}_{i\frac{m}{b},\frac{l}{2}} \left( \frac{1}{2}br^2 \right) \\
  \psi_{2+}(r) = \left(\frac{4}{b}\right)^\frac{l+1}{2} \frac{\Gamma\left( \frac{l}{2} + \frac{1}{2} +  i\frac{m}{b} \right)}{\Gamma(l+1) \Gamma\left(- \frac{l}{2} + \frac{1}{2} +  i\frac{m}{b} \right)} \frac{1}{\sqrt{r}} \mathcal{M}_{i\frac{m}{b}, \frac{l}{2}} \left(\frac{1}{2}br^2 \right)
 \end{eqnarray}
 where, $\mathcal{W}_{a,b}(z)$ and $\mathcal{M}_{a,b}(z)$ are the Whittaker functions \cite{Hans, Taylor, Abad}. Now, one can identify the following assymptotic behavour.
 
  \begin{itemize}
 \item{ \begin{equation}
 \psi_{1+}(r) \sim \left(\frac{4}{b}\right)^\frac{l+1}{2} \frac{1}{\sqrt{r}} \left(\frac{1}{2}br^2 \right)^{i\frac{m}{b}} e^{-\frac{b}{4}r^2} , \;\; \mbox{as}\;\; r\rightarrow \infty
 \end{equation}
  Therefore, $l$ independently it goes to zero at infinity.}
 
 \item{ \begin{equation}
 \psi_{2+}(r) \sim \left(\frac{4}{b}\right)^\frac{l+1}{2} \frac{\Gamma\left( \frac{l}{2} + \frac{1}{2} +  i\frac{m}{b} \right)}{\Gamma(l+1) \Gamma\left(- \frac{l}{2} + \frac{1}{2} +  i\frac{m}{b} \right)} \frac{1}{\sqrt{r}} e^{\frac{1}{4}br^2} \left( -\frac{b}{2}r^2 \right)^{-i\frac{m}{b}} ,\;\; \mbox{as}\;\; r\rightarrow \infty
 \end{equation}
  So, it diverges at infinity and this diverging nature is independent of $l$. So, it is not square integrable in $(0,\infty)$. Therefore $\psi_{2+}(r)$ is not a feasible solution. }
 
 \item{
   \begin{equation}
    \psi_{1+}(r)\sim r^{-l-\frac{1}{2}} \left[ r + \beta_1 r^3 -\beta_2 r^5 + r^{2l+1} \left( \beta_3 +\beta_4 r^2 - \beta_5 r^4 +..... \right)  \ \right],\;\; \mbox{as}\;\; r\rightarrow 0
    \end{equation}
     Where, $\beta_1 = -\frac{b}{4}\alpha_4 + \alpha_5,\; \beta_2 = \frac{b}{4}\alpha_5 ,\; \beta_3 = \alpha_1 \alpha_2  ,\; \beta_4 = \alpha_1 \alpha_3 -\frac{b}{4}\alpha_1\alpha_2 ,\; \beta_5 = \frac{1}{4} \alpha_1 \alpha_3$. \\
     Where, $\alpha_1 = e^{2l\pi}                               ,\;\;\;
     \alpha_2 = \frac{(\frac{b}{2})^{\frac{1}{2}+ \frac{l}{2}}\Gamma(-l)}{ \Gamma(\frac{1}{2}+ \frac{l}{2}-\frac{mi}{b})}                                       ,\;\;\;\;\;\;
     \alpha_3 = \frac{(\frac{b}{2})^{\frac{1}{2}+ \frac{l}{2}}\Gamma(-l)}{4\Gamma(l+1) \Gamma(\frac{1}{2}+ \frac{l}{2}-\frac{mi}{b})} (-2mi+b+bl)                                       \\
     \alpha_4 =  \frac{(2b)^{-\frac{1}{2}+ \frac{l}{2}}\Gamma(l)}{\Gamma(\frac{1}{2}+ \frac{l}{2}-\frac{mi}{b})}                                      ,\;\;\;\;\;\;\;\;\;
     \alpha_5 = \frac{(2b)^{-\frac{1}{2}+ \frac{l}{2}}\Gamma(l)}{4(l-1) \Gamma(\frac{1}{2}+ \frac{l}{2}-\frac{mi}{b})} (2mi-b+bl) $.\\\\
     Clearly, $\forall l \geq 1$, this diverges and it only converges for the case $l=0$.
      }
 \end{itemize}
 
 Therefore, we have seen that (27) has only one normalizable solution for $l=0$. Similar situation occurs for (28). Therefore, for $l \neq 0$, $n_+=n_-=0$, i.e, $H_l$ is self-adjoint operator in $L^2((0,\infty),rdr)$. Whereas, for $l=0$, the deficiency index $n_+=n_-=1$, that means infinitely many self-adjoint extensions are possible by an one parameter unitary maping. In particular,
 \begin{equation}
 \mbox{domain}\;\;  h_0^{\theta} = \{ \phi+ c(\psi_{1+}- e^{i\theta} \psi_{1-}) ; \phi \in \mbox{dom}\; \bar{h_0},\;  c \in \mathcal{C},\; \theta \in [0,2\pi] \}.
 \end{equation}
 Then one can conclude the following theorems. 
 
 \begin{thm}
 $\hat{H}_l$ is self adjoint for $l\neq 0$.
 \end{thm}
 
 \begin{thm}
 There exist self-adjoint extensions characterized by one parameter unitary map for the hermitian operator $\hat{H}_0$.
 \end{thm}
 The proof of the above theorems are trivially followed from the previous arguments. 
 This situation is similar as that of Coulomb potential problem in two dimension \cite{Cesar} .
 Actually, this may be general feature for the spherically symmetric potential, though further study is required.

 \section{Conclusions}
 Thus we have seen that, the case $l=0$ is special in the sense that it has infinite number of self
 adjoint extension. The case of the boundary condition which describes the Landau level wave function for
 anyon, merely an special case of our obtained class of self-adjoint extension. It is worth noting that,
 the formalism described here was also used previously to obtain the self-adjoint extension for Coulomb
 potential. We have obtained the similar results for our case. Therefore, one can hope that the formalism
 is far more general in the sense, it can be used for all possible spherically symmetric potential.
 and the extension of this for all possible spherically symmetric potential will be an interesting one. 

 \section{Acknowledgement}
 J P Saha is grateful to DST, Govt. of India for financial support under DST-PURSE scheme.

\end{document}